\PassOptionsToPackage{numbers,super,square,comma,sort&compress}{natbib}
\documentclass[pdflatex,twocolumn,iicol]{sn-jnl}

\usepackage{graphicx}
\usepackage{amsmath,amssymb,amsfonts}
\usepackage{amsthm}
\usepackage{bm}
\usepackage{url}
\usepackage[english]{babel}
\usepackage{multirow}
\usepackage{enumerate}
\usepackage{cases}
\usepackage{dsfont}
\usepackage{color,soul}
\usepackage{tcolorbox}
\usepackage{booktabs}
\usepackage{array}
\usepackage{tabularx}
\usepackage{hyperref}
\usepackage{natbib}

\renewcommand{\arraystretch}{1.5}
\title{Low-Altitude Wireless Networks: The Next Horizon of Wireless Infrastructure}
\author{\fnm{Yuanhao} \sur{Cui}}
\author{\fnm{Jiali} \sur{Nie}}
\author{\fnm{Weijie} \sur{Yuan}}
\author{\fnm{Fan} \sur{Liu}}
\author{\fnm{Ziye} \sur{Jia}}
\author{\fnm{Jie} \sur{Xu}}
\author{\fnm{Zhiyong} \sur{Feng}}
\author{\fnm{Mohamed-Slim} \sur{Alouini}}
\abstract{Low-altitude airspace, roughly defined as the region up to 3000 meters above ground level, is envisioned as a new spatial domain for daily human and machine activities. This article introduces the concept of the Low-Altitude Wireless Network (LAWN), which represents a paradigm shift from the current ground-based communication-only network to a three-dimensional (3D) multifunctional network. We analyze the key driving forces, network architecture, and limiting factors of LAWN, with a particular focus on the tight integration of communication, sensing, and control in highly dynamic airspace environments. By establishing the coupling between airspace capacity and wireless channel capacity, we reveal the intrinsic limits of airspace management and identify the fundamental challenges and opportunities associated with its evolution.}

\begin{document}
\maketitle


The success of wireless networks, as one of the most impactful technological advances in human history, has significantly revolutionized humankind's daily life and led to worldwide infrastructure expansion. Today, more than 70\% of terrestrial areas and 90\% of the global population are supported by wireless infrastructure offering uninterrupted communication service, yet these systems generally maintain a ground-based approach in both design and execution~\cite{LinCSM2021,you2021towards}. This omission reflects a historical focus on human-centric connectivity, overlooking the imminent demand for three-dimensional (3D) wireless services to support emerging activities in low-altitude airspace, e.g., electric vertical takeoff and landing (eVTOL) aircraft and unmanned aerial vehicles (UAVs)~\cite{zaid2023evtol,liu2022integrated,yuan2025ground}.

\begin{figure*}[t]
    \centering
    \includegraphics[width=1\textwidth]{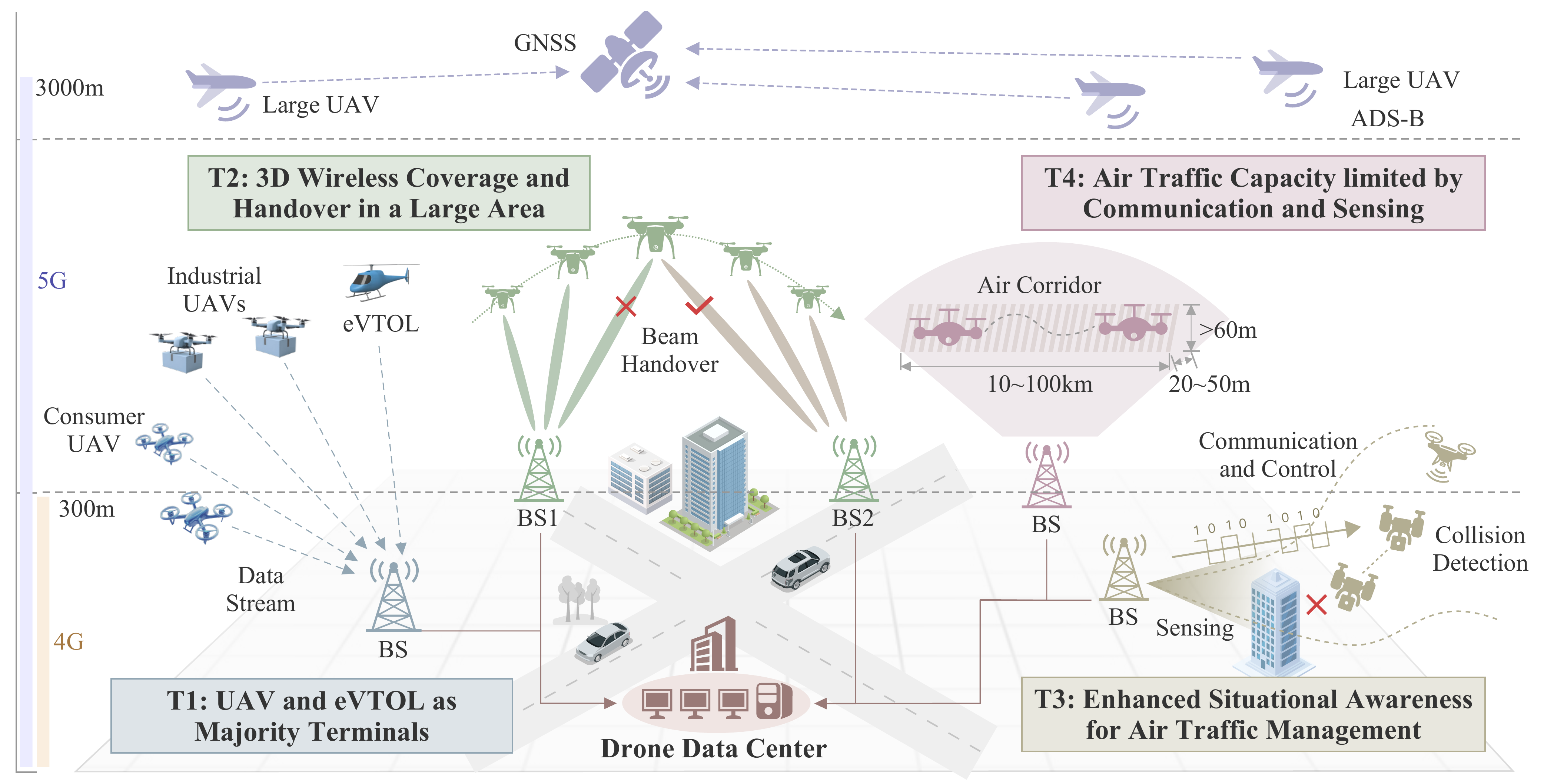}
    \caption{\textbf{The digitized low-altitude skyway and its evolutionary trends.} This conceptual figure illustrates the multi-layer 3D architecture of LAWN as a structured skyway that integrates ISAC-enabled ground infrastructure, data centers, and heterogeneous airborne terminals to provide connectivity, high-resolution sensing, and autonomous traffic orchestration. The four main drivers highlighted in the figure are (T1) UAVs and eVTOLs as major aerial terminals with massive uplink data streams, (T2) 3D wireless coverage and handover across a large service area, (T3) enhanced situational awareness for air traffic management, and (T4) air traffic capacity constraints that are jointly shaped by communication and sensing performance within structured air corridors.}
    \label{fig:lawn_layers}
\end{figure*}

However, the widespread deployment of UAVs and the rapid maturation of eVTOL platforms now amplify this urgency. For example, worldwide regulatory agencies are expected to grant more than 100 airworthiness certifications by 2030, paving the way for the commercialization of low-altitude transportation systems~\cite{ChinaD}. Driven by demand for drone delivery and automated cargo transport, industry projections estimate that the low-altitude logistics market could reach 60 billion dollars by 2035. Consequently, terrestrial cellular networks are not well suited to meet the communication and sensing demands created by the projected growth of low-altitude vehicle traffic over the next decade~\cite{tang2025cooperative}.

Therefore, we are witnessing a paradigm shift in wireless infrastructure from a ground-based communication-only network to a skyward multifunctional network, namely the low-altitude wireless network (LAWN)~\cite{wu2025low}. Beyond providing reliable communication connectivity, LAWN is envisioned as the wireless infrastructure underpinning low-altitude air transport through a unified operational framework that combines surveillance, navigation, and traffic orchestration. This transformation would fundamentally redefine cellular networks, shifting them from passive utilities to active enablers of a digitized 3D ``skyway'', in which the infrastructure autonomously supports low-altitude transportation safety and efficiency~\cite{yuan2025ground,huang2024potential}.

Closing this gap between existing cellular networks and LAWN requires innovation at both the physical and architectural levels. At the physical level, networks require integrated sensing and communication (ISAC) to merge environmental monitoring with data transmission, enabling base stations (BSs) to track aircraft while maintaining connectivity~\cite{9606831}. At the architectural level, interoperability with digitized airspace management platforms is non-negotiable. By integrating ISAC-enabled BSs with air traffic management systems, wireless networks can dynamically map airspace, construct air corridors, enforce collision-avoidance protocols, and automate routing, thereby transforming raw bandwidth into structured mobility governance~\cite{JiangCMAG2025}. LAWN is therefore expected to become an important component of next-generation wireless infrastructure.

\section{Drivers and Vision of LAWN}
Low-altitude airspace, roughly defined as the region up to 3000 meters above ground level, as shown in Fig.~\ref{fig:lawn_layers}, is poised to become a critical frontier for both human and machine activity in the coming decade. This space will support a diverse set of applications, from parcel-delivery drones and urban air taxis to precision-agriculture UAVs and public-safety surveillance. This exponential growth in airborne connectivity and sensing services demands a paradigm shift in network design.

\subsection{Communication and Sensing Trends}

\textbf{Trend 1: UAVs and eVTOLs as Major Aerial Terminals:} The proliferation of drones and eVTOL aircraft introduces a new class of heterogeneous wireless terminals operating in the sky~\cite{zaid2023evtol,namuduri2022advanced}. By 2030, the number of cellular-connected UAVs is projected to reach tens of millions globally, rivaling smartphones in scale. These airborne devices require enormous data traffic volumes and ultra-reliable low-latency links for safety-critical control at scale~\cite{YuIoTJ2023,mohsan2023unmanned,zhang2025energy}. For instance, a small fleet of drones can produce roughly 150~TB of data per day, so tens of millions of UAVs could collectively generate hundreds of petabytes of data daily. Unlike relatively homogeneous mobile phones, UAVs vary widely in size, altitude, and mission profile, resulting in highly diverse sensing and communication requirements~\cite{11098638}. This heterogeneity leads to complex connectivity challenges such as intermittent signal blockage (e.g., by buildings or terrain), extremely frequent handovers due to high mobility, and the need for adaptive protocols that maintain control links and target tracking for fast-moving aerial vehicles~\cite{mohsan2022towards}.

\textbf{Trend 2: 3D Wireless Coverage and Handover in a Large Area:} Providing seamless and robust coverage in three dimensions is a fundamental requirement for LAWN. In this setting, BSs (or other access nodes) must reliably serve users at various altitudes, including near-ground, rooftop, and several-hundred-meter levels, without signal gaps or weak zones~\cite{10125057,10542290}. For instance, a delivery drone or eVTOL should maintain network connectivity throughout its flight path, which requires seamless handover among BSs in both the vertical and horizontal planes. Therefore, additional rooftop BSs that fill vertical coverage gaps are often required in LAWN design, ensuring that low-altitude vehicles remain connected throughout the airspace corridor~\cite{shayea2022handover}.

\textbf{Trend 3: Enhanced Situational Awareness for Air Traffic Management:} Real-time situational awareness, high-precision terrain information, and nearby weather awareness are crucial for flying vehicles, which in turn require LAWN to function as a reliable situational-awareness infrastructure~\cite{9464266,10328607}. Flying vehicles must not operate ``blind''; instead, the network and associated management systems should continuously monitor the surroundings of each aircraft to prevent accidents. In addition, the feasibility of dynamic airspace management, route planning, and collision avoidance in low-altitude airspace is highly dependent on the sensing and communication capacity of the infrastructure. For example, NASA envisions predefined routes or corridors and ``dynamic geofences'' around no-fly zones, with networked warnings to ensure that a drone does not stray into a dangerous area. The width and height of air corridors are therefore tightly constrained by the sensing and communication precision available from the infrastructure. In this context, LAWN would effectively augment onboard sensors with widely distributed, high-resolution, all-weather, day-and-night ``eyes and ears'' on the ground~\cite{hamissi2023survey}.

\textbf{Trend 4: Air Traffic Capacity Constrained by Communication and Sensing:} Air traffic capacity is jointly limited by self-positioning, sensing, and control precision~\cite{bulusu2017capacity}. For example, a drone travelling at 15~m/s can cover 1 to 2~m in 0.1~s, which may be the difference between stopping short of another drone and colliding with it. Moreover, drones flying near the ground are subject to urban-canyon effects~\cite{1177161}, in which signals such as GPS and radio links can be blocked or reflected by tall structures. A drone flying between high-rise buildings may lose line of sight to GPS satellites or BSs; even moderate-height buildings or terrain can cause intermittent GPS reception, which further limits the usable airspace and its traffic capacity~\cite{10049809}. On the other hand, sensing precision also constrains flying speed. For example, a small quadcopter cruising slowly can hold its position within a few centimeters and navigate around objects carefully, whereas at speeds above 20~m/s the position error or overshoot may become orders of magnitude larger, increasing collision risk.

\subsection{Design Objectives of LAWN}

\begin{figure*}
    \centering
    \includegraphics[width=1\textwidth]{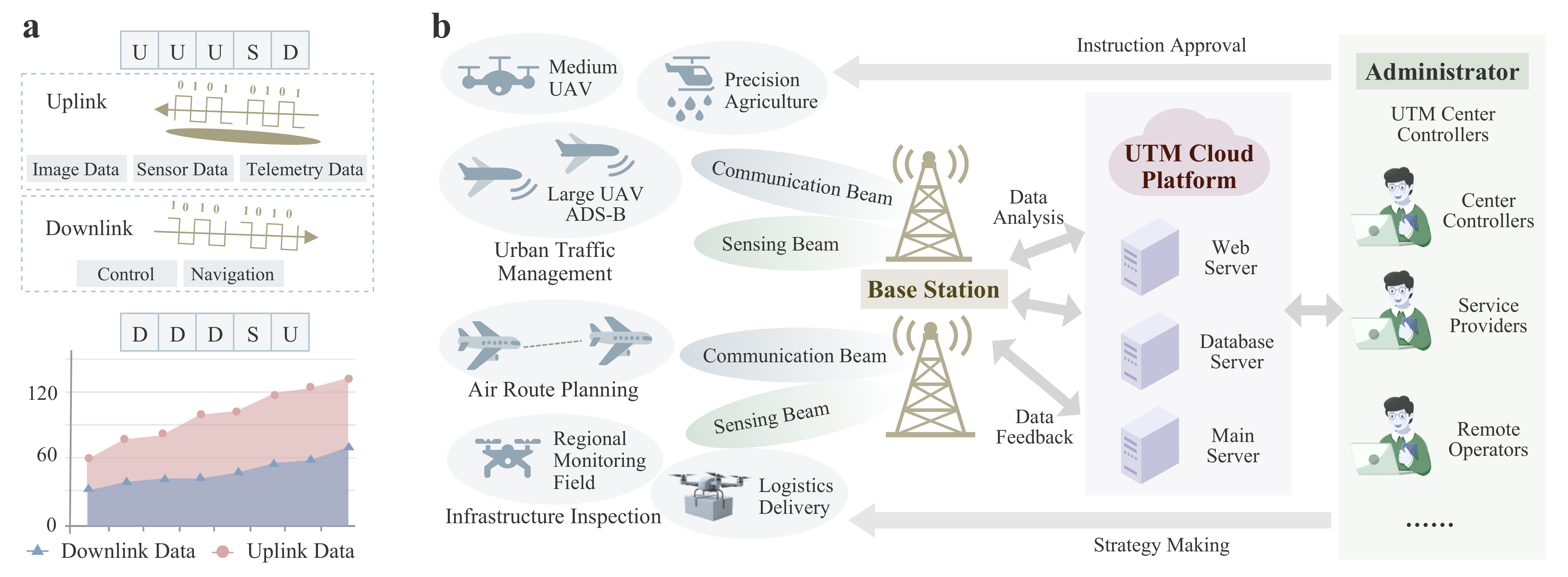}
    \caption{\textbf{System architecture of LAWN with tightly integrated communication, sensing, and unmanned traffic management (UTM).} \textbf{a,} Comparison between a conventional downlink-dominant terrestrial frame structure (``DDDSU'') and an uplink-oriented LAWN frame structure (``UUUSD''), reflecting heavy image, sensor, and telemetry uploads together with lighter control and navigation downlinks. \textbf{b,} ISAC-enabled BSs provide communication and sensing beams for mission services such as regional monitoring, infrastructure inspection, logistics delivery, precision agriculture, and air-route planning, while a cloud UTM platform coordinates data analysis, feedback, strategy making, and instruction approval for administrators, controllers, service providers, and remote operators.}
    \label{fig:lawn_layers1}
\end{figure*}

The overarching goal of LAWN is to serve as the wireless infrastructure underpinning low-altitude air transport, thereby ensuring the efficiency and safety of flying vehicles such as UAVs and eVTOLs. To meet this objective, LAWN must go beyond traditional cellular networks and seamlessly integrate communication, sensing, and air traffic management functions, as illustrated in Fig.~\ref{fig:lawn_layers1}b. Accordingly, LAWN should provide not only reliable and low-latency connectivity but also become an integral part of digitized airspace~\cite{huang2024potential}. In this vision, each BS can act as a basic airspace-management unit, providing high-resolution situational-awareness services to surrounding aerial vehicles.

\section{Network Architecture and Limiting Factors}
The transition from ground-centric cellular systems to a fully 3D LAWN architecture explicitly reflects the vertical diversity of low-altitude airspace. To provide both reliable connectivity and real-time traffic management, LAWN must be decomposed into altitude-specific layers, each with its own mix of terrestrial, aerial, and satellite links, as shown in Fig.~\ref{fig:lawn_layers}. In this section, we first discuss the layered network architecture of LAWN and then examine its key limiting factors.

\subsection{Altitude-Specific Layered Operation}
\subsubsection{Below 300 m}
Low-altitude airspace below $300$~m is typically used for urban air mobility and local drone services. Representative applications include eVTOL air taxis within cities and last-mile delivery drones dropping off packages in neighborhoods~\cite{ha2023last}. Public-safety and infrastructure-monitoring missions are also common in this layer, such as inspection drones for utilities like power lines and railroads.

Terrestrial BSs are expected to play a major role in connectivity for aerial vehicles below $300$~m~\cite{9392776}. At these heights, drones often maintain line of sight to multiple cell towers, which yields strong received signals but also increases interference from neighboring cells. Therefore, proper interference-management techniques are required for LAWN evolution, such as adjusting antenna tilt or using sectors optimized for sky coverage. With these enhancements, an urban eVTOL or delivery drone can be served by LAWN for telemetry, control, and payload communications. In addition, the network must interface with UTM systems by providing real-time tracking and identification of each drone over the cellular link~\cite{prevot2016uas}. Overall, a terrestrial-network-centric architecture is expected to handle the dense connectivity demand, provided that it is tightly integrated with airspace traffic management for access control and safety~\cite{honghong2022risk}.

\subsubsection{Between 300 m and 1,000 m}
The $300$--$1000$~m altitude layer primarily supports medium-range aerial operations, including cross-city cargo drones and inter-city eVTOL air taxis. UAVs operating in this range can cover distances of tens of kilometers, serving regional logistics, medical-supply delivery, and inter-district transportation~\cite{geraci2022integrating, geraci2022will}.

In this intermediate airspace, vehicles fly above building heights but below traditional commercial airplanes, typically within uncontrolled airspace except near airports. Therefore, to ensure safe and efficient operations, LAWN-based UTM systems must evolve to handle higher flight speeds, extended mission durations, and seamless oversight across broader regions. Maintaining continuous, wide-area connectivity becomes particularly challenging at these altitudes, as UAVs frequently move beyond the coverage of standard terrestrial networks. While terrestrial BSs can still provide connectivity along parts of these routes, a purpose-designed network is often required to support long-distance airborne operations. 
One possible approach is deploying high-power BSs on tall towers to extend cell range for aerial users. However, ground BSs may be sparse and cannot cover the entire 3D space, and UAV relays may be employed to fill this gap. 

\subsubsection{Above 1,000 m}
Above $1000$~m, low-altitude UAV networks begin to intersect with the lower bounds of traditional aviation airspace. This category includes large UAVs for surveillance and remote sensing~\cite{huang2018uav,guo2019drone,mathews2023fundamental} and high-flying drone swarms for wide-area monitoring~\cite{asaamoning2021drone, saffre2023wild}. For example, military or border-surveillance drones often cruise at altitudes of a few kilometers to gain a broad view, and environmental-monitoring UAVs may operate at similar heights to cover large regions.

The operational airspace above $1000$~m is often regulated, which requires UAVs to obtain flight clearance and continuous tracking similar to that of manned aircraft, especially if they share air corridors with piloted planes. Thus, interoperability with the conventional air traffic management system is a key requirement for LAWN. At these altitudes, aerial and satellite communications become the primary connectivity solutions. LEO satellite networks (such as Starlink) can offer broadband data, enabling drones to stay connected to the internet or control centers even over oceans or remote regions~\cite{liu2018space,zhang2025uav}. In addition, experimental high-altitude long-endurance (HALE) drones, originally designed for Earth observation, are now being explored as airborne communication platforms.

\subsection{Key Limiting Factors for LAWN}
Despite rapid progress, several factors still limit the deployment and scaling of LAWN.
\subsubsection{Propagation \& Channel Characteristics}
The wireless channel in low-altitude airspace can vary from dense multipath near the ground to nearly free-space propagation at higher altitudes. Near the ground in urban areas ($<300$~m), signals are obstructed by buildings, leading to shadowing and multipath fading that can cause link unreliability and dead zones~\cite{khawaja2019survey}. High mobility can also introduce pronounced Doppler shifts~\cite{sandri2023implementation}. These physical-layer issues demand advanced link adaptation, beamforming, and potentially integrated sensing to continuously track and maintain robust links in 3D space~\cite{giordani2020non}.
\subsubsection{Network Layer Challenges}
LAWN must handle extreme mobility in three dimensions. A drone can travel through multiple cell sectors in minutes, leading to very frequent handovers. Without intelligent handover management, UAVs may drop connections or suffer high latency during handoff. Moreover, current network routing and cell association algorithms, optimized for quasi-static ground users, may perform poorly for aerial nodes. Supporting seamless mobility requires multi-cell connectivity that allows drones to connect to multiple towers simultaneously or predictive handover techniques based on trajectory \cite{rinaldi2020non}. The network layer also faces scalability concerns as millions of airborne devices demand routing, addressing, and QoS management in the sky, significantly increasing load on network infrastructure.

\subsubsection{Asymmetric Downlink/Uplink Data Traffic} 
Fig.~\ref{fig:lawn_layers1}a highlights a fundamental asymmetry between uplink (UAV-to-BS) and downlink (BS-to-UAV) traffic, with LAWNs exhibiting significantly higher service demands in the uplink. Aerial nodes typically generate large volumes of data that must be transmitted from air to ground, including high-definition video streams and real-time sensor data~\cite{zhou2021real}. Conversely, downlink transmissions from ground to air mostly consist of brief control commands and navigation updates, which are comparatively lightweight. This imbalance exists consistently across all altitude layers. Accordingly, substantial changes are required in resource allocation, antenna configuration, and protocol design relative to ground-based networks~\cite{zhang20196g}. Antenna systems require vertical beamforming adaptations or upward-oriented coverage patterns to better receive and manage aerial transmissions.

\subsubsection{Spectrum Management} 
Communication links in LAWN often share spectrum with ground networks~\cite{lee2023feasibility}. Using existing cellular bands for drones can therefore cause mutual interference. In particular, drone signals transmitted from altitude can travel farther and potentially interfere with distant cell sites. The issue of spectrum allocation for UAVs is only beginning to be addressed by regulators. Careful spectrum planning, dynamic frequency coordination, and possibly new frequency bands (e.g., millimeter-wave for high-capacity drone-to-drone links or optical communications at higher altitudes) will be needed to accommodate LAWN growth.

\subsubsection{Regulations} 
The deployment of LAWN is tightly coupled with evolving regulations~\cite{azari2022evolution}. Aviation authorities impose strict rules on unmanned aircraft operations. For example, routine flights are commonly limited to altitudes around $120$~m. Expanding LAWN operations to medium or higher altitudes will require new airspace classifications. In addition to allocating suitable frequency bands, regulators may need to constrain transmit power to minimize interference. Furthermore, privacy and cybersecurity requirements will significantly shape network design, including encryption protocols and authentication mechanisms.

\section{Digitized Airspace Governance: Modeling, Structuring, and Co-Design}

\begin{figure*}
    \centering
    \includegraphics[width=0.98\textwidth]{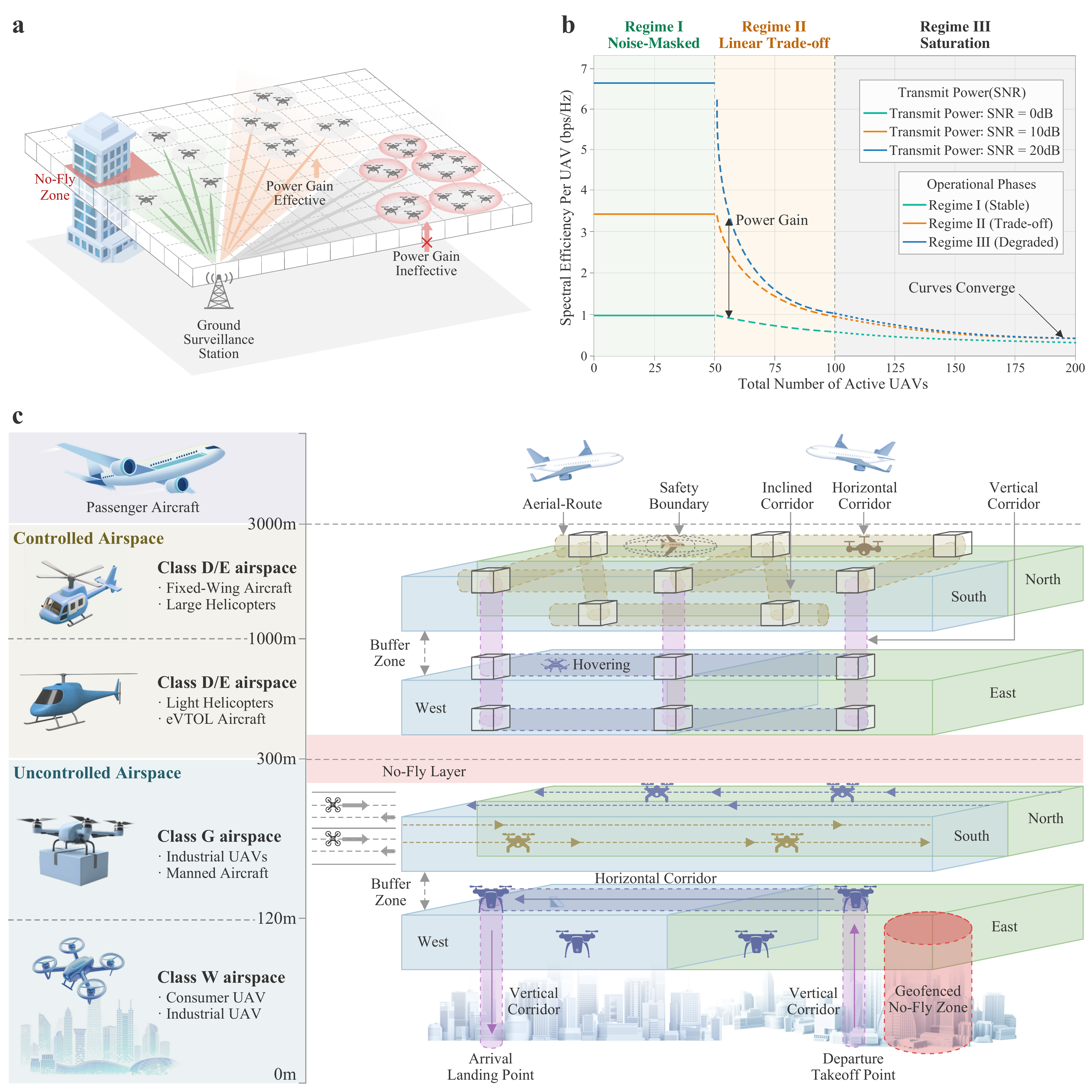}
    \caption{\textbf{Air-corridor-aware airspace structuring and the airspace--channel coupling limit in LAWN.} \textbf{a,} Beam-occupancy illustration in a discretized airspace: as concurrent UAV density increases, some users remain in high-gain sectors whereas others become interference-limited near no-fly constraints. \textbf{b,} Spectral efficiency per UAV versus total active UAVs under different SNR conditions, revealing three operating regimes (noise-masked, linear trade-off, and saturation/degraded) and convergence at high load. The shaded regime boundaries are illustrated using the low-SNR case ($\rho=0~\mathrm{dB}$) for visual clarity and should be interpreted as conceptual operating regions rather than exact thresholds for all SNR curves. \textbf{c,} Multi-layer corridor design from 0 to 3000\,m, including controlled and uncontrolled airspace classes, vertical, horizontal, and inclined corridors, buffer zones, and geofenced no-fly areas for heterogeneous users such as consumer and industrial UAVs, eVTOLs, helicopters, and passenger aircraft.}
    \label{fig:aircorridor}
\end{figure*}

Airspace management is foundational to scaling low-altitude applications because it encompasses the systematic organization and dynamic control of airspace to ensure the safe, efficient, and equitable integration of users such as UAVs and eVTOLs~\cite{li2022traffic}. The design of structured air corridors is particularly important because it enhances operational safety and efficiency by segregating traffic flows, minimizing cross-path conflicts, and enabling predictable routing. In addition, conflict-resolution mechanisms must be embedded within such corridors, leveraging real-time sensing and communications to support collision avoidance and automated geofencing against incursions by unauthorized objects or non-cooperative flights~\cite{seo2017collision,yasin2020unmanned,rezaee2024comprehensive}.

\subsection{Airspace-Channel Coupling and Capacity Limits}

The fundamental challenge of LAWN lies in the intrinsic coupling between airspace management and wireless channel capacity \cite{huang2024potential}. Unlike terrestrial cellular systems, where users are largely confined to a two-dimensional plane, low-altitude airspace introduces a 3D, mobile, and congestion-prone environment, in which spatial conflicts directly translate into communication interference \cite{shoewu2022uav,luo2025toward}. To establish a unified analytical foundation, it is necessary to model low-altitude airspace capacity and channel capacity within a common framework \cite{Bulusu2017CapacityEF}.

\subsubsection{Airspace and Channel Capacity Definition}
The 3D airspace $\mathcal{D}$ is discretized into $N$ uniform grid cells, indexed by $\mathcal{G}=\{1,\ldots,N\}$. The global traffic state is represented by a position matrix $\mathbf{X} \in \{0,1\}^{K \times N}$, where $K$ denotes the total number of active UAVs. The airspace capacity ($C_{\text{air}}$) is defined as the macroscopic traffic volume supported by the system, corresponding to the $L_1$-norm of the traffic state:
\begin{equation}
\begin{aligned}
C_{\text{air}}(\mathbf{X}) \triangleq \sum_{k=1}^{K}\sum_{n=1}^{N} X_{k,n} &= K, \\
\text{s.t.} \quad 
\mathbf{X}\mathbf{1}_N &= \mathbf{1}_K, \\
\mathbf{X}^T\mathbf{1}_K &\le \mathbf{c}_{\text{geo}}.
\end{aligned}
\end{equation}
where the constraint $\mathbf{X}\mathbf{1}_N = \mathbf{1}_K$ ensures that each UAV occupies exactly one grid cell, while $\mathbf{X}^T\mathbf{1}_K \le \mathbf{c}_{\text{geo}}$ imposes the geometric occupancy limit of each cell \cite{huang2024potential}. 

The channel capacity characterizes the achievable spectral efficiency for user $k$ given the spatial configuration $\mathbf{X}$ and precoding $\mathbf{W}$:
\begin{equation}
C_{\text{chan},k}(\mathbf{X}) = \log_2(1 + \gamma_k(\mathbf{W}, \mathbf{X}))
\end{equation}
where $\gamma_k$ is the Signal-to-Interference-plus-Noise Ratio (SINR). Under a practical regime of coherent intra-beam interference, the SINR is governed by the Beam Occupancy Level ($\mu_k$), which counts the active UAVs falling within the interference subspace of user $k$:
\begin{equation}
\gamma_k(\mathbf{X}) \approx \frac{\rho}{(\mu_k(\mathbf{X}) - 1)\rho + 1}
\end{equation}
where $\rho \triangleq P_{\text{rx}}/\sigma^2$ denotes the raw Signal-to-Noise Ratio (SNR).

\subsubsection{Phase Transition of Airspace–Channel Coupling}

The relationship between $C_{\text{air}}$ and $C_{\text{chan}}$ exhibits a nonlinear phase transition governed by the interaction between thermal noise and spatial interference. By identifying the critical capacity threshold $C_{\text{crit}} = L(1 + 1/\rho)$, where $L$ is the number of orthogonal beams, the airspace capacity can be divided into three distinct operational regimes, as shown in Fig.~\ref{fig:aircorridor}a,b:
\begin{itemize}
\item \textbf{The Noise-Masked Regime: } Occurring when $C_{\text{air}} \le L$, the system is noise-limited. In this sparse state, $\mu_k = 1$, and $C_{\text{chan}}$ is decoupled from $C_{\text{air}}$, bounded only by transmit power.
\item \textbf{The Linear Trade-off Regime:} In the region $L < C_{\text{air}} \le C_{\text{crit}}$, beam collisions occur, but the interference power is comparable to the noise floor. A linear trade-off exists where joint power-space optimization is most effective.
\item \textbf{The Saturation Regime:} When $C_{\text{air}} \gg C_{\text{crit}}$, the system becomes strictly interference-limited. The channel capacity degrades inversely with airspace density: $C_{\text{chan}} \approx \log_2(1 + \frac{1}{C_{\text{air}}/L - 1})$. In this regime, power control is ineffective, and spatial separation or deconfliction becomes the dominant strategy.
\end{itemize}

\subsubsection{QoS-Constrained Capacity Boundary}
To guarantee safe operation, the system must maintain a minimum spectral-efficiency requirement $R_{\text{min}}$ (in bit/s/Hz). The maximum allowable airspace capacity is thus theoretically bounded by:
\begin{equation}
C_{\text{air}}^{\text{max}} \le L \left( 1 + \frac{1}{2^{R_{\text{min}}} - 1} - \frac{1}{\rho} \right)
\end{equation}

This derivation provides a theoretical foundation for LAWN airspace management, highlighting that airspace throughput is fundamentally limited by beam occupancy, which is in turn shaped by sensing precision, interference management, and available communication resources.

\subsection{Structured Air Corridors: Organizing 3D Mobility}

Air corridors are reserved 3D volumes designed to structure aerial traffic, replacing free-flight regimes with predefined pathways and thereby enhancing both operational safety and traffic efficiency~\cite{aggarwal2020path,muna2021air}. Unlike uncontrolled airspace, air corridors enforce predetermined flight trajectories to minimize collision risks, reduce communication and sensing overhead, and prevent incursions into restricted zones~\cite{muna2021air}.

\subsubsection{Geometric Layering and Spatial Discretization}
The fundamental concept of the air corridor involves discretizing and layering the continuous 3D low-altitude airspace based on altitude, orientation, and mission attributes. As illustrated in Fig.~\ref{fig:aircorridor}c, this architectural shift replaces free-flight regimes with a multi-layered model in which the airspace is partitioned into altitude-specific corridor units to enhance operational safety and traffic efficiency. In a typical implementation, the airspace is divided into three functional layers: the top and bottom layers serve as directional conduits, accommodating southbound/northbound and eastbound/westbound traffic, respectively, thereby segregating flows to minimize mid-air cross-path conflicts. A specialized middle layer acts as a transition zone, providing a reserved volume for aircraft to hover or adjust altitude when changing direction, thereby supporting fluid and coordinated traffic orchestration. The width and height of these corridors are tightly constrained by the sensing and communication precision provided by LAWN infrastructure~\cite{wang2025evaluating}. By aligning these physical dimensions with network capability, the system enforces predetermined flight trajectories and reduces the likelihood of incursions into restricted zones.

\subsubsection{Communication-Space Resource Mapping}
Each air corridor implicitly corresponds to a finite communication resource budget, including the number of available beams, tolerable interference levels, and minimum communication-rate constraints. By confining UAV flight paths within designated corridors, the infrastructure can effectively control the spatial density of UAVs. This structuring limits the beam-occupancy level $\mu_k$, preventing the system from exceeding the critical threshold $C_{\text{crit}}$ and entering a performance-collapse regime. Furthermore, the corridors are strategically aligned with the high-gain regions of BS beamforming patterns to support uninterrupted telemetry and control links. This mapping between 3D airspace and communication resources is therefore a prerequisite for ultra-reliable aerial operations~\cite{karimi2024optimizing}.

\subsubsection{Mission-Specific Architectures}
LAWN implements a multi-tier strategy for the design of dedicated air corridors tailored to specific mission requirements and operational priorities. For example, medical transport corridors may prioritize shortest-path routing with emergency right-of-way privileges, whereas cargo delivery corridors could focus on noise mitigation and industrial-zone alignment, following rigorous risk stratification. This mission-specific optimization is further bolstered by dynamic geofencing capabilities, which allow the network to enforce automated boundaries around no-fly zones \cite{haddad2021traffic}. By leveraging networked warnings and real-time situational awareness, the infrastructure ensures that UAVs do not stray into restricted or hazardous areas, thereby transforming the airspace into a managed, task-aware mobility layer \cite{munir2022situational}.


\begin{figure*}
    \centering
    \includegraphics[width=0.98\textwidth]{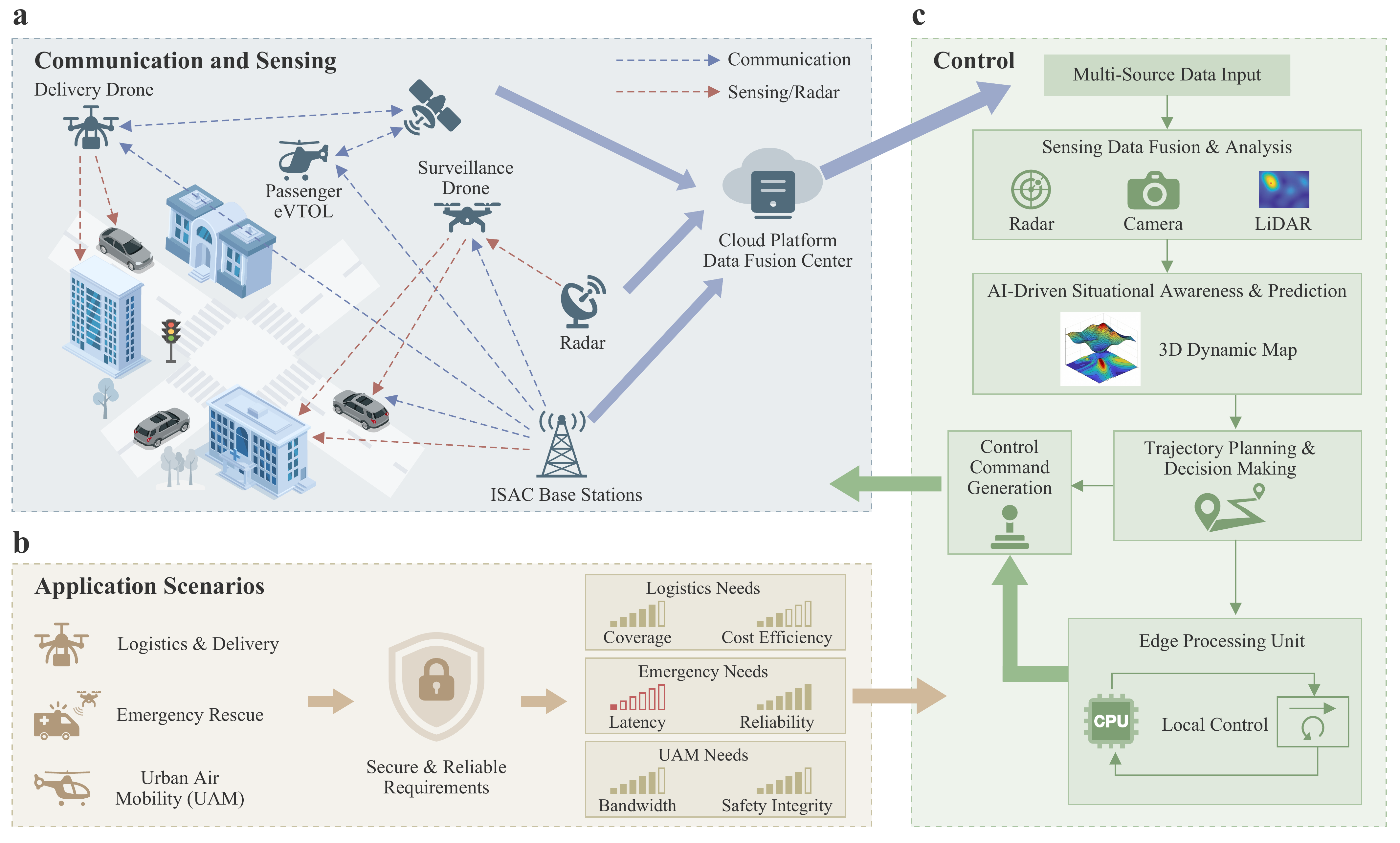}
    \caption{\textbf{CSC co-design framework for closed-loop LAWN operations.} \textbf{a,} Delivery drones, passenger eVTOLs, surveillance UAVs, radar, and ISAC BSs exchange communication and sensing links, and multi-source observations are fused at a cloud data platform. \textbf{b,} Representative applications (logistics, emergency rescue, and urban air mobility) map to differentiated requirement vectors in coverage, latency, reliability, bandwidth, cost efficiency, and safety integrity. \textbf{c,} Fused radar/camera/LiDAR data drive AI situational awareness, 3D dynamic mapping, trajectory planning, and control-command generation, while edge processing units provide local feedback control for real-time and safety-critical actuation.}
    \label{fig:CSC}
\end{figure*}

\subsection{Physics-Aware CSC Co-Design: Closing the Loop}

\subsubsection{Communication Model}
Most existing systems treat communication, sensing, and control as loosely coupled functional modules, which struggle to operate effectively in highly dynamic LAWN environments~\cite{jin2026advancing}. Conventional approaches often rely on idealized or quasi-static communication assumptions, thereby overlooking the stochastic latency, packet loss, and time-varying reliability inherent in wireless channels~\cite{wang2023qos}. In the context of closed-loop control over wireless links, the actual execution of the control command is conditioned on successful packet delivery. Specifically, the discrete-time dynamics of the UAV can be formulated as:
\begin{equation}
\mathbf{q}_{t+1} = \mathbf{A}\mathbf{q}_t + \alpha_t \mathbf{B}\mathbf{u}_t + \mathbf{n}_t,
\end{equation}
where $\mathbf{q}_t$ denotes the system state vector, $\mathbf{A}$ is the state transition matrix characterizing the inherent physical evolution of the plant, $\mathbf{B}$ is the control input matrix, $\mathbf{u}_t$ is the control command generated by the BS based on the latest estimates, and $\mathbf{n}_t$ represents the process noise. The binary variable $\alpha_t \in \{0, 1\}$ characterizes the communication packet loss at time slot $t$, which follows a Bernoulli distribution:
\begin{equation}
\alpha_t = 
\begin{cases} 
1, & \text{with probability } p_t(\mathbf{w}_t), \\
0, & \text{with probability } 1-p_t(\mathbf{w}_t),
\end{cases}
\end{equation}
where $p_t(\mathbf{w}_t)$ is the packet success probability (PSP). Unlike static models, the PSP in our framework is intrinsically coupled with the beamforming design $\mathbf{w}_t$. To capture the nonlinear relationship between the Signal-to-Interference-plus-Noise Ratio (SINR) and link reliability, we employ a sigmoid-based mapping function:
\begin{equation}
p_t(\mathbf{w}_t) = \frac{1}{1 + \exp\left(-a\left(\text{SINR}_t(\mathbf{w}_t) - \gamma_{\text{th}}\right)\right)},
\end{equation}
where $a$ is the steepness coefficient and $\gamma_{\text{th}}$ represents the target SINR threshold. Any failure of the beamforming gain to maintain $\text{SINR}_t$ above the threshold $\gamma_{\text{th}}$ results in a sharp escalation of the packet loss probability. For an unstable system, such communication-induced outages can lead to the exponential growth of the state error, potentially violating the topological stability requirements discussed in the sequel.

\subsubsection{Sensing Model}
From the sensing perspective, the BS simultaneously performs target sensing by processing the echoes of the transmitted signal. Let $\mathbf{q}_t$ be the true kinematic state of the UAV. The BS obtains the state estimate $\hat{\mathbf{q}}_t$ through echo processing, which is defined as:
\begin{equation}
\hat{\mathbf{q}}_t = \mathbf{q}_t - \mathbf{e}_t,
\end{equation}
where $\mathbf{e}_t$ denotes the state estimation error. In our framework, this error is modeled as a zero-mean Gaussian random variable with covariance $\mathbf{\Sigma}_t = \mathbb{E}[\mathbf{e}_t \mathbf{e}_t^T]$. Based on estimation theory, the lower bound of this covariance is characterized by the Cramér-Rao Bound (CRB), which is the inverse of the Fisher Information Matrix (FIM). Specifically, when estimating the spatial parameter $\theta$ (e.g., Angle-of-Arrival), the FIM $\mathbf{J}(\theta)$ is intrinsically coupled with the beamforming design $\mathbf{w}_t$. For a system with $L$ snapshots and $N_r$ receive antennas, the CRB for angle estimation is given by:

\begin{equation}
\text{CRB}(\theta) = \frac{\sigma_s^2}{2 L |\beta|^2 N_r \cdot |\dot{\mathbf{a}}^H(\theta) \mathbf{w}_t|^2},
\end{equation}
where $\sigma_s^2$ is the noise variance, $L$ represents the number of snapshots used for sensing accumulation, $N_r$ is the number of receive antennas, $\beta$ is the complex channel gain, and $\dot{\mathbf{a}}(\theta)$ represents the derivative of the steering vector $\mathbf{a}(\theta)$ with respect to $\theta$. This bound reveals a fundamental trade-off, indicating that minimizing sensing uncertainty requires the beamforming vector $\mathbf{w}_t$ to be aligned with the derivative of the steering vector $\dot{\mathbf{a}}(\theta)$ rather than with the steering vector itself.

The sensing uncertainty directly propagates into the control domain. By applying the state-feedback control law $\mathbf{u}_t = -\mathbf{K} \hat{\mathbf{q}}_t$, the control input becomes:
\begin{equation}
\mathbf{u}_t = -\mathbf{K}(\mathbf{q}_t - \mathbf{e}_t) = -\mathbf{K}\mathbf{q}_t + \mathbf{K}\mathbf{e}_t.
\label{CSC}
\end{equation}
where $\mathbf{K}$ represents the state-feedback gain matrix. Then, the control law can be expanded as:
\begin{subequations} \label{eq:dynamics_all}
\begin{align}
    \mathbf{q}_{t+1} &= \mathbf{A} \mathbf{q}_t + \alpha_t \mathbf{B} \mathbf{u}_t + \mathbf{n}_t, \label{eq:open_loop} \\
    &= (\mathbf{A} - \alpha_t \mathbf{B}\mathbf{K})\mathbf{q}_t + \alpha_t \mathbf{B}\mathbf{K}\mathbf{e}_t + \mathbf{n}_t, \label{eq:closed_loop}
\end{align}
\end{subequations}

To ensure the stability of the closed-loop system under stochastic packet loss and sensing uncertainty, we define a quadratic Lyapunov function $V(\mathbf{q}_t) = \mathbf{q}_t^T \mathbf{P} \mathbf{q}_t$, where $\mathbf{P} \succ \mathbf{0}$ is a positive definite matrix \cite{polycarpou1993robust}. The stability of the system is guaranteed if the one-step conditional expectation of the Lyapunov function satisfies the following drift constraint \cite{freeman2008robust}:
\begin{equation} \label{eq:lyapunov_constraint}
    \mathbb{E}[V(\mathbf{q}_{t+1}) \mid \mathbf{q}_t] - \eta V(\mathbf{q}_t) \le 0,
\end{equation}
where $\eta \in (0, 1)$ denotes the target decay rate. This formulation defines the safety boundary of the system, where $\mathbb{E}[\cdot]$ is intrinsically coupled with both the packet success probability $p(\mathbf{w}_t)$ and the sensing-error covariance represented through $\text{CRB}(\mathbf{w}_t)$. To satisfy~(\ref{eq:lyapunov_constraint}), the beamforming vector $\mathbf{w}_t$ must be judiciously designed to navigate the fundamental trade-off between enhancing communication reliability $p(\mathbf{w}_t)$ and suppressing the state-estimation error $\mathbf{e}_t$.

\subsubsection{Entropy-Based CSC Co-Design}
To navigate the complexities of the emerging low-altitude economy, LAWN must transcend the conventional siloed design paradigm and evolve toward a deeply integrated communication, sensing, and control (CSC) co-design architecture \cite{11045436}. As illustrated in Fig.\ref{fig:CSC}, this architecture establishes a closed-loop ecosystem where multi-source data streams from ISAC-enabled BSs, terrestrial radars, and UAV platforms are aggregated at cloud-edge platforms. These fused data streams drive AI-powered situational awareness modules, enabling precise trajectory planning and real-time navigation command generation \cite{xu2023uav}. Furthermore, the CSC framework supports adaptive resource scheduling to accommodate the heterogeneous mission requirements inherent in LAWN \cite{liang2024sensing}. These range from the wide-area coverage and moderate latency needs of logistics delivery to the stringent safety-integrity and ultra-reliable low-latency constraints of urban air mobility.

Unlike traditional wireless systems where reliability is primarily defined by the ability to decode signals against noise, reliability in CSC co-design is intrinsically linked to the physical stability of the UAV. According to the data rate theorem, a linear unstable system is stabilizable only if the feedback communication rate $R(\mathbf{w}_t)$ exceeds the system's inherent uncertainty expansion rate, known as the topological entropy \cite{MineroTAC2009}:
\begin{equation}
\mathcal{H}(\mathbf{A}) = \sum_{|\lambda_i(\mathbf{A})| > 1} \log_2 |\lambda_i(\mathbf{A})| ,
\end{equation}
where $\lambda_i(\mathbf{A})$ are the unstable eigenvalues of the state transition matrix. This necessitates an entropy reduction constraint on the beamforming design:
\begin{equation}
\text{SINR}(\mathbf{w}_t) \ge \underbrace{2^{\frac{\mathcal{H}(\mathbf{A})}{B_{\rm W}}} - 1}_{\gamma_{\text{critical}}},
\end{equation}

where ${B_{\rm W}}$ is the communication bandwidth and $\gamma_{\text{critical}}$ delineates the absolute survival threshold. Any misalignment that causes $\text{SINR}(\mathbf{w}_t) < \gamma_{\text{critical}}$ renders the system unstabilizable, as the information flow becomes insufficient to neutralize physical entropy production~\cite{5276196}. Beyond mere survival, the CSC architecture must guarantee stability and state convergence. Consequently, the CSC co-design problem transitions into a survival-constrained joint optimization~\cite{8681425}:

\begin{subequations} \label{eq:CSC_Optimization}
\begin{align}
\min_{\mathbf{w}_t} \quad & \Vert \mathbf{w}_t \Vert^2 \tag{P1} \\
\text{s.t.} \quad & \text{SINR}(\mathbf{w}_t) \ge \gamma_{\text{critical}}, \label{con:entropy} \\
& \mathbb{E}[V(\mathbf{q}_{t+1}) \mid \mathbf{q}_t] \le \eta V(\mathbf{q}_t), \label{con:lyapunov} 
\end{align}
\end{subequations}
where \eqref{con:entropy} ensures topological stabilizability, while \eqref{con:lyapunov} enforces kinematic convergence. By aligning communication reliability and sensing precision with the entropy-based stability requirement, the CSC architecture ensures that the infrastructure dynamically prioritizes information fidelity and latency based on the instantaneous kinematic risk of aerial platforms~\cite{6740858,zheng2023control,scheuvens2021state}. This approach transforms LAWN from a ``best-effort'' data pipe into a physics-aware resilient network capable of supporting safety-critical urban air mobility.

\section{Challenges and Opportunities in LAWN}
The differences between LAWN and traditional cellular networks are summarized in Table~\ref{tab:lawn_vs_terrestrial}. To fully harness the potential of LAWN, critical technical challenges spanning both the physical and network layers must be addressed~\cite{zeng2016wireless}. These challenges not only underscore the complexity of deploying robust LAWN systems but also present unique research opportunities across communication and sensing protocols, resource management, and network scalability. In this section, we identify key obstacles in physical-layer enhancement, sensing services, and network and airspace management, and propose practical directions to advance the feasibility and performance of LAWN architectures.

\subsection{Physical-Layer Enhancement in LAWN}
As established in prior sections, the physical layer of LAWN exhibits intrinsic operational contrasts with terrestrial systems, such as the asymmetric downlink/uplink data traffic. This inversion of traditional traffic dynamics, where downlink services typically dominate, stems from LAWN-specific use cases, such as UAV-based data collection and aggregation or real-time telemetry, necessitating tailored physical-layer design strategies \cite{7470933}. Therefore, several physical-layer enhancements are required in LAWN.
\subsubsection{Dedicated Frame Structure Design}
The conventional 5G NR frame structure (e.g., ``DDDSU'') is no longer suited to LAWNs because of their inherent uplink-centric demands. In this configuration, as shown in Fig.~\ref{fig:lawn_layers1}a, ``D'' denotes downlink slots, ``U'' represents uplink slots, and ``S'' signifies a flexible special slot comprising mixed downlink and uplink symbols together with a guard period (GP), which prioritizes downlink-heavy traffic in typical terrestrial networks~\cite{3gpp2020study}. In contrast, LAWNs require dynamic slot allocation that favors uplink capacity. This mismatch underscores the need for adaptive frame structures tailored to LAWNs' asymmetric traffic patterns.
\subsubsection{Task-Oriented Uplink Transmission}
In LAWN applications such as environmental monitoring, infrastructure inspection, and media broadcasting, UAVs collect substantial sensor data that require real-time transmission to BSs under minimal distortion. To optimize this process, semantic communication techniques may be employed to compress data in a content-aware and task-oriented manner, ensuring that only important and relevant information is delivered~\cite{9398576}. For example, during infrastructure inspections of power lines or pipelines, UAVs can use semantic compression to focus on transmitting data related to potential anomalies or defects. This approach reduces bandwidth usage and accelerates issue identification.
\subsubsection{Multi-Functional URLLC Downlink Transmission}
Unlike terrestrial networks, LAWN downlinks operate as multifunctional platforms rather than simple data pipes, converging sensing, communication, and control into a unified framework. This integration equips BSs with continuous, ubiquitous, and all-weather airspace surveillance via downlink wireless sensing, enabling early detection of unauthorized UAV intrusions~\cite{10539181}. Simultaneously, BSs leverage sensing-derived spatial information to dynamically adjust UAV flight paths through downlink navigation commands, which is a critical capability in GNSS-denied environments. To realize real-time, seamless, and robust control of large fleets of UAVs, downlink data must be delivered at very low latency (typically below 10~ms), where URLLC techniques may play a vital role.

\begin{table*}[ht]
\centering
\caption{Requirement Metrics: LAWN vs. Traditional Cellular Networks}
\label{tab:lawn_vs_terrestrial}
\begingroup
\renewcommand{\arraystretch}{1.2}
\small
\begin{tabularx}{\textwidth}{>{\raggedright\arraybackslash}p{3.7cm} >{\raggedright\arraybackslash}X >{\raggedright\arraybackslash}p{7.7cm}}
\toprule
\textbf{Category} & \textbf{Traditional Cellular Network} & \textbf{Low-Altitude Wireless Network (LAWN)} \\
\midrule
Primary Service Targets & Smartphones, IoT devices & UAVs, eVTOLs, aerial sensors, airborne radars \\
Coverage Dimension & 2D ground coverage & 3D integrated ground-air coverage (0--3000 m) \\
Typical Mobility & Ground-based ($<$120 km/h) & Mission-dependent 3D aerial mobility (tens to hundreds of km/h) \\
Uplink/Downlink Balance & Downlink-dominant ($>$70\%) & Uplink-dominant ($>$60\%) due to telemetry and sensing \\
Frequency Bands & Sub-6 GHz, partial mmWave & Sub-6 GHz, mmWave, potential THz \\
Latency Requirement & 10--100 ms (eMBB) & $<$10 ms (control); millisecond-level for sensing and safety-critical functions \\
Throughput Requirement & Downlink: 10--100 Mbps average & Uplink: 50--200 Mbps average; up to Gbps level \\
Reliability Requirement & $\sim$99.9\% (standard QoS) & $>$99.999\% (critical UAV control) \\
Communication Range & $\sim$200--500 m (urban macro cell) & Several km in LoS; dynamic 3D coverage \\
Sensing Capability & Weak or absent & Integrated sensing and communication (ISAC) \\
Interference Pattern & Horizontal inter-cell interference & Vertical cross-domain interference (UAV $\leftrightarrow$ UE) \\
QoS Management & RRM, slicing, HARQ, MCS & Task-aware dynamic scheduling, trajectory-based QoS \\
Deployment Topology & Fixed macro/micro BSs & Hybrid deployment with flying relays and fixed BSs \\
Sensing Accuracy & $\sim$10 meters (GNSS-based) & Meter-level to sub-meter-level (target detection and tracking) \\
Handover Complexity & Horizontal handover (2D) & Frequent 3D handovers, CoMP support \\
User/Device Density & $\sim$1 million devices/km$^2$ & Fewer users but high dynamics and spatial complexity \\
Energy Constraints & Moderate (smartphones/IoT) & High (energy-constrained UAVs) \\
Service Scenarios & Video streaming, IoT, VoLTE & Remote control, real-time sensing, HD video uplink \\
\bottomrule
\end{tabularx}
\endgroup
\end{table*}

\subsection{Sensing and ISAC}
ISAC, which integrates radio sensing and wireless communication into a unified system~\cite{liu2022integrated}, plays a foundational role in enabling the full capabilities of LAWN. This convergence equips wireless infrastructure with real-time sensing, dynamic target tracking, and situational-awareness capabilities in low-altitude airspace.
\subsubsection{Sensing Coverage Enhancement}
Conventional terrestrial networks are conceived for 2D ground coverage, typically offering reliable communication service only up to roughly 100~m in vertical height from sub-6~GHz macro BSs. However, aerial platforms in LAWN can operate from near-ground altitudes to around 3000~m, and potentially beyond in some suburban or rural scenarios. This vertical disparity creates sensing-coverage gaps that must be addressed. One effective strategy is to expand the vertical beamforming angle of the transmit antenna array. Recent experiments within the framework of 5G-Advanced (5G-A) networks have demonstrated that increasing the vertical beamforming angle to $65^\circ$ can substantially enhance integrated ground-to-air sensing and communication performance~\cite{bayesteh2022integrated}. Specifically, a well-designed monostatic ISAC-enabled active antenna unit (AAU) operating at sub-6~GHz has been shown to support reliable sensing at altitudes up to 600~m. Additionally, cooperative sensing and beam coordination among multiple BSs can be employed to realize seamless sensing in complex urban environments.

\subsubsection{Tradeoff between UAV Sensing and Terrestrial User Communication}
One of the key challenges in LAWN arises from integrating wireless sensing functionality into the network infrastructure, which introduces an inherent trade-off between UAV sensing and communication service delivery to ground users~\cite{10908560}. To achieve the desired sensing performance, certain BSs must adjust the tilt angles of their antenna arrays to steer spatial beams toward UAVs~\cite{lyu2022joint,meng2023uav}. However, such beam steering often diminishes radiated signal power toward the ground, resulting in a substantial decline in communication rates for terrestrial users. A potential solution is to deploy dedicated BSs at elevated positions that exclusively provide ISAC services to low-altitude UAVs. While effective, this approach incurs significant costs in terms of infrastructure deployment and network planning. As a more practical alternative, advanced ISAC signal processing and resource-allocation strategies can be implemented in a ``best-effort'' manner using existing BSs~\cite{zhang2021enabling}. These strategies aim to minimize sensing loss functions, such as the mean squared error (MSE) in UAV localization and tracking, while simultaneously upholding the quality-of-service (QoS) requirements of ground users~\cite{zhang2023sensing,dong2023joint}.

\subsubsection{Low-Slow-Small Target Detection and Recognition}
Low-altitude airspace is increasingly populated with low, slow, and small (LSS) aerial targets, such as consumer drones, birds, and micrometeorological probes, which pose challenges for both civilian applications and airspace security. Most such targets are characterized by low radar cross-section (RCS, often below --20~dBsm), slow and irregular motion (e.g., 5--20~m/s), and limited RF emissions, making them difficult to detect or recognize using conventional radar platforms~\cite{khawaja2025survey}. Tests conducted by the UK Civil Aviation Authority showed that small quadcopters, such as the DJI Phantom series with an RCS as low as --23~dBsm, could not be reliably tracked beyond 500~m using standard airport-surveillance radar. Tackling this issue necessitates close collaboration among large numbers of BSs in LAWN. By exploiting the spatial diversity of a dense BS deployment, the network can correlate faint signal anomalies across multiple observation points and thereby improve detectability.

\subsubsection{Air-to-Ground Sensing}
Air-to-ground sensing serves as a critical enabler for situational awareness in LAWN, especially as UAVs increasingly function not only as communication relays but also as mobile sensing platforms~\cite{10529184}. Unlike terrestrial sensing systems constrained by obstructions and horizon limits, aerial platforms benefit from broader LoS coverage and flexible mobility, enabling real-time environmental mapping from elevated vantage points. A key advantage lies in their ability to augment terrestrial perception through cross-perspective fusion~\cite{tong2023multi}. In a coordinated LAWN setting, UAVs operating at altitudes of 100--200~m can help identify blind spots in ground-based sensing, such as alleyways or overpasses. For instance, when aerial LiDAR depth maps are fused with mmWave radar returns from ground BSs, it becomes feasible to reconstruct partial occlusions and estimate pedestrian trajectories with higher confidence. Nevertheless, the altitude and speed of aerial platforms cause rapid geometric changes in observation angle, resulting in significant variation in elevation angle and slant range, which in turn leads to complex backscattering effects and geometric distortion. Airborne mmWave radar mounted on low-flying UAVs can detect ground vehicles in urban canyons with sub-meter accuracy, but performance degrades significantly when targets are shielded by metallic structures or dense foliage.

\section{Outlook}

In this article, we have presented a vision of low-altitude wireless networks as a new wireless infrastructure for digitized airspace. Starting from the driving trends and design objectives of LAWN, we discussed its layered network architecture, key limiting factors, and the intrinsic coupling among airspace capacity, wireless channel capacity, and closed-loop control stability. Compared with traditional cellular networks, LAWN represents a paradigm shift from communication-oriented connectivity to a 3D multifunctional infrastructure that integrates communication, sensing, and control. In this vision, BSs are expected to serve not only as communication access points, but also as basic airspace-management units that support situational awareness, traffic orchestration, and safety-critical control. These new requirements and applications also introduce a range of research challenges and opportunities, spanning physical-layer enhancement, sensing service provision, network resource management, and airspace governance. Addressing these challenges will be essential for transforming LAWN from a conceptual extension of cellular networks into a reliable air-transport infrastructure for future low-altitude economies.

\bibliographystyle{bst/sn-nature}
\bibliography{ref,sn-bibliography}

\end{document}